\documentclass[a4paper]{article}

\usepackage[pages=all, color=black, position={current page.south}, placement=bottom, scale=1, opacity=1, vshift=5mm]{background}
\SetBgContents{
}      

\usepackage[margin=1in]{geometry} 

\usepackage{amsmath}
\usepackage{amsthm}
\usepackage{amssymb}

\usepackage[utf8]{inputenc}
\usepackage{hyperref}
\hypersetup{
	unicode,
	pdfauthor={Author One, Author Two, Author Three},
	pdftitle={A simple article template},
	pdfsubject={A simple article template},
	pdfkeywords={article, template, simple},
	pdfproducer={LaTeX},
	pdfcreator={pdflatex}
}

\usepackage[sort&compress,numbers,square]{natbib}
\theoremstyle{plain}

\theoremstyle{definition}

\usepackage{graphicx, color}
\graphicspath{{fig/}}

\usepackage{algorithm, algpseudocode} 
\usepackage{mathrsfs} 

\usepackage{lipsum}

\title{Ladder operators approach to representation classification problem for Jordan-Schwinger image of su(2) algebra}
\author{Tushavin G.V.$^1$ 
\thanks{This work was supported by the Ministry of Science and Education of the Russian Federation (Passport No. 2019-0903).} 
\and Trifanov A.I.$^1$ \and Zaitseva E.V.$^1$}

\date{
	$^1$ITMO University \\ 
}

\begin{document}
	\maketitle
	
\begin{abstract}
	The eigenvalues of the complete commuting set of self-adjoint operators determine the classifi-
	cation of states. We construct a classification for the image of the Jordan–Schwinger mapping of the su(2)
	algebra. We use the ladder operator approach to construct a canonical basis of irreducible representations
	and define the self-adjoint operators of the complete commuting set. \\
	
\end{abstract}

\tableofcontents

\section{Introduction}
\label{sec:intro}
	
The study of dynamics of some quantum systems can be reduced to the study of the dynamic group of the Hamiltonian. 
Generators of the dynamical group form an algebra. The structures of invariant spaces of the algebra and the
group are similar. Eigenvalues of self-adjoint operators of the complete commuting set are used to the state 
classification. The ladder operator approach used to build the complete set and obtain eigenbasis. In articles  \cite{1,2,3,4} 
ladder operators are constructed for different algebras, which are obtained in consequence of modification of 
quantum harmonic oscillator model. In our work, we have formulated a general approach to the analysis of such systems.

The Lie algebra of the dynamic group of the Hamiltonian of the quantum harmonic oscillator model is a             ()
Heisenberg-Weyl algebra \cite{5, 6} ~--- $w(1)$. Generators of this algebra are hermitian-conjugate boson 
creation/annihilation operators ~--- $a$ and $a^\dagger$ obey the following commutation relations
\begin{equation}\label{algebra_w1}
	\left[ a, \, a^\dagger \right] = \hat{I}, \quad 
	[a, \, \hat{I}] = 0 = [a^\dagger, \, \hat{I}].
\end{equation}
Here $\hat{I}$ ~--- the identity operator of the algebra $w(1)$.
By introducing a particle number operator $\hat{N} = a^\dagger a$ the mentioned Hamiltonian may be expressed as
\begin{equation}
	\hat{H} = \hbar \omega \left(\hat{N} + \frac{1}{2}\right).
\end{equation}
The complete commuting set of operators for this Hamiltonian contain only one operator $\hat{N}$, which spectrum determines 
the observed energy levels. Operators $a$ and $a^\dagger$ are \textit{ladder operators} for the operator $\hat{N}$. They 
satisfy the commutation relations
\begin{equation}\label{ladder_w1}
	\left[\hat{N}, a^\dagger \right] = a^\dagger, \quad \left[\hat{N}, a \right] = -a.
\end{equation}
Action of ladder operators $a$ and $a^\dagger$ translates an eigenvector of operator $\hat{N}$ into another eigenvector
\begin{equation}\label{ladder_act}
\begin{matrix}
	\hat{N} \left| n \right> = n \left| n \right>, \quad
	\hat{N} (a^\dagger \left| n \right> ) = 
	a^\dagger (\hat{N} + \hat{I}) \left| n \right> = (n + 1) (a^\dagger \left| n \right> ),
	\\
	\hat{N} (a \left| n \right> ) = 
	a(\hat{N} - \hat{I}) \left| n \right> = (n - 1) (a \left| n \right> ), \quad
	a \left| 0 \right> = 0 = \hat{N}\left| 0 \right>,
\end{matrix}
\end{equation}
The annihilation operator $a$ (unlike the creation operator $a^\dagger$) has a non-trivial kernel 
corresponding to the vacuum state of the quantum oscillator. The corresponding eigenvector 
$\left| 0 \right>$ is usually called \textit{vacuum vector}. Thus, the spectrum of operator 
$\hat{N}$ consists of integer non-negative numbers $\mathbb{N}\cup \{0\}$, and an arbitrary 
eigenvector can be obtained by the action of ladder operators on any particular eigenvector, e.g. 
the vacuum vector. In the canonical basis of the eigenvectors $\{ \left| n \right> \}$ the 
operators $a$ and $a^\dagger$ have the form
\begin{equation}
	a^\dagger \left| n \right> = \sqrt{n + 1}  \left| n + 1 \right>, \quad
	a \left| n + 1 \right> = \sqrt{n + 1}  \left| n \right>, \quad
	a \left| 0 \right> = 0,
\end{equation}
and vector $ \left| n \right> $ is expressed as
\begin{equation}
	\left| n \right> = \frac{1}{\sqrt{n!}} (a^\dagger)^n \left| 0 \right>.
\end{equation}

In this way, dynamics of multidimensional harmonic oscillator may be described by an algebra which generators 
are represented through the bosonic polynomials resulting from Jordan-Schwinger mapping \cite{7, 8}
of generator matrices into $w(1)^{\otimes m}$ for certain $m$:
\begin{equation}\label{def_Jordan}
	X = (x_{ij}) \mapsto \breve{X} = \sum_{i,j = 1}^{m}x_{i,j}a_i^\dagger a_j, \quad [\breve{X},\,\breve{Y}] = \breve{[X, \, Y]}.
\end{equation}
The image of the identity matrix is the total particle number operator 
\begin{equation}
	\breve{I} = N = \sum\limits_{\mu = 1}^{m} a_{\mu}^{\dagger} a_{\mu} = 
		\sum\limits_{\mu = 1}^{m} N_{\mu}.
\end{equation}
In our paper \cite{10} we consider an image of the algebra $su(2)$ \cite{9} represented by the operators
$N, \, J_z, \, J_+, \, J_-$, which are expressed through bosonic operators $a_i, \, a_i^\dagger$ 
by the Jordan-Schwinger mapping of generator matrices of the irreducible representation of dimension $(2s+1)$ of $su(2)$ algebra \cite{8}:
\begin{equation}\label{def_su2_Jordan}
		J_z = \sum\limits_{\mu = -s}^{s} \mu a_{\mu}^{\dagger} a_{\mu},\quad
		J_+ = \sum\limits_{\mu = -s}^{\mu = s - 1} \sqrt{(s + \mu + 1)(s - \mu)} a_{\mu + 1}^{\dagger} a_ {\mu} = (J_-)^{\dagger}.
\end{equation}
We denote this algebra $su^j(2)$. Bosonic operators for each degree of freedom obey the following commutation relations:
\begin{equation}
	[a_i, a_j] = [a_i^\dag, a_j^\dag] = 0, \quad [a_i, a_j^\dag] = \delta_{ij}.
\end{equation}
The Fock basis, defined by eigenvalues of particle number operators for each degree of freedom, 
is complete and consists of vectors of the form $ \left| n_{-s}, n_{-s+1}, \ldots, n_s \right>$.

The Jordan-Schwinger mapping is a Lie algebras homomorphism, thus the matrix and their images obey the same 
commutation relations. The operators $J_z, \, J_+, \, J_-$ represent the generators of $su(2)$ algebra and satisfy 
the corresponding commutative relations
\begin{equation}
	[J_z, \, J_\pm] = \pm J_\pm, \quad [J_+, \, J_-] = 2J_z.
\end{equation}
In $su(2)$ the Casimir operator commuting with all generators exists. By Schur's lemma, in each space of 
irreducible representation such an operator is proportional to the identity operator. Recall, that the image 
of the unit matrix in the Jordan-Schwinger mapping is operator $N$. Hence
\begin{equation}
	[N, \, J_z] = 0, \quad [N, \, J_\pm] = 0.
\end{equation}
The Casimir operator $J^2$ for generators $J_z, \, J_+, \, J_-$ is defined as follows
\begin{equation}
	J^2 = J_z^2 + \frac{1}{2}(J_+ J_- + J_- J_+),
\end{equation}
and canonical basis for each irreducible representation has a standard form
\begin{gather*}
	\begin{matrix}
		J_z \left| j, \, j_z \right> = j_z \left| j, \, j_z \right>, 
		\quad
		J^2 \left| j, \, j_z \right> = j(j+1) \left| j, \, j_z \right>,\\
		J_+ \left| j, \, j_z = j \right>  = 0, \quad
		J_- \left| j, \, j_z = -j \right>  = 0,\\
		J_+  \left| j, \, j_z \right> = \sqrt{(j - j_z)(j + j_z + 1)} \left| j, \, j_z + 1 \right>, \\
		J_-  \left| j, \, j_z + 1 \right> = \sqrt{(j - j_z)(j + j_z + 1)} \left| j, \, j_z \right>.
	\end{matrix}
\end{gather*}

The commuting set of operators $\left\{N; J^2, J_z\right\}$ is complete when $s = \frac{1}{2}$ and $s = 1$.  
In these cases, the eigenvalues of operators $N, J^2, J_z$ uniquely determine basis vectors 
$\left|n; \, j, \,j_z \right>$. In other cases within a fixed eigenvalue $n$ of the operator $N$, the
eigenvalues $j(j+1)$ of the operator $J^2$ are nontrivially degenerate.  Note that if $s$ is a non-negative 
integer, then $j$ is also a non-negative integer. The arbitrary Fock vector will be an eigenvector for 
the operators $\left\{N; J_z\right\}$, but not for the operator $J^2$.

The aim of our work is to augment the existing commutative set $N;\, J^2, \, J_z$ to a complete one. 
In our paper we propose a method for constructing generalized ladder operators, which are used for 
classification and construction of the canonical basis.
		
\section{Ladder operators: motivation}

\subsection{Generalized ladder operators}

Let us consider the self-adjoint operator $H=H^\dagger$ from some finitely generated algebra $\mathcal{A}$. 
We will call an operator $p^\dagger$ a \textit{right ladder operator} (hereafter, RLO) if there exists a 
nonzero selfadjoint operator $P=P^\dagger \neq 0$ from $\mathcal{A}$ commuting with $H$, such that one of 
the following equivalent commutation relations is satisfied

\begin{equation}\label{def_ladderRight}
	[H, \, p^\dagger] = p^\dagger P  \quad \text{or} \quad Hp^\dagger = p^\dagger (P + H).
\end{equation} 
The expression conjugated to \eqref{def_ladderRight} is the definition of the \textit{left ladder operator} (LLO)
\begin{equation}\label{def_ladderLeft}
[p, \, H] = P p.
\end{equation} 
For RLO $p^\dagger$ we will call the operator $P$ a \textit{right function} in the case when the operator $P$ is represented as a function $P(H, H_1, \ldots, H_n)$ of the commuting set of   self-adjoint operators $H, \, H_1, \, ,\ldots, H_n$.

For an arbitrary polynomial of the operator $H$ the operator $p$ is a ladder operator. In view of the bilinearity of the commutator it suffices to show that for any degree of $H$ the following property holds

\begin{equation}\label{exp_HnPdag}
	[H^n, \, p^\dagger] =  p^\dagger((H+P)^n-H^n) \quad \text{or} \quad H^n p^\dagger = p^\dagger (H+P)^n.
\end{equation}
\textbf{Proof.}
To prove this statement its anought to use the recurrent property 
\begin{equation}
  [H^n, \, p^\dagger] = [H, \, p^\dagger]H^{n-1} + H[H^{n-1}, p^\dagger] = 
  p^\dagger P H^{n-1} + [H^{n-1}, p^\dagger](H + P),
\end{equation}
which is completed by applying the method of mathematical induction, where the base of induction is the definition of the ladder operator. 

It is worth to note that for any operator $A$ the following expression holds:
\begin{equation}
	[H p^\dag A] = p^\dag [H+P, A] + p^\dag AP,
\end{equation}
and one can show that multiplying the RLO by the self-adjoint operator $A \in \mathcal{A}$ which is commute 
with operator $(H + P)$ is again the RLO of operator $H$:
\begin{equation}\label{comm_HPA}
    [H, \, p^\dag A] = p^\dag AP.
\end{equation}

\subsection{Ladder operators construction}

Let us consider the self-adjoint operator $H$ and the set of operators $\{T_\mu\}_{\mu = 1}^n$ which have the 
following properties: there exist a family of mutually commuting operators $\{\alpha_{\mu\eta}\}$ 
such that
\begin{equation}\label{act_HTmu}
	[H, \, T_\eta] = \sum_{\mu=1}^{n}T_\mu \alpha_{\mu\eta}, \quad 
	\alpha_{\mu \eta}^{\dag} = \alpha_{\mu \eta}, \quad [\alpha_{\mu \eta}, H] = 0.
\end{equation}
We are looking for a nontrivial set of self-adjoint operators $\sigma_\eta = {\sigma_\eta}^\dagger$ 
which commute with $H$ and $\{\alpha_{\mu\eta}\}$ and the operator $\sum_{\eta=1}^{n} T_\eta \sigma_\eta$ is the RLO for $H$ again:
\begin{equation*}
    [H, \, \sum_{\eta=1}^{n} T_\eta \sigma_\eta] = \sum_{\eta=1}^{n} T_\eta \sigma_\eta P, \quad [P, \, \sigma_\eta] = 0.    
\end{equation*}
Substituting \eqref{act_HTmu} into the previous expression, we obtain the following equation
\begin{equation}\label{eq_spec}
	\sum_{\mu=1}^{n} T_\mu (\sum_{\eta = 1}^{n}\alpha_{\mu\eta} \sigma_\eta - \sigma_\mu P) = 0,
\end{equation}
which may be represented in matrix form
\begin{equation}\label{eq_eigenmat}
	\left(
	\begin{matrix}
		T_1 & T_2 & \ldots & T_n
	\end{matrix}
	\right)
	\left(
	A-P
	\right)
	\left(
	\begin{matrix}
		\sigma_1 \\
		\sigma_2 \\
		\ldots \\
		\sigma_n \\
		
	\end{matrix}
	\right) = 0,
\end{equation}
where we use $(A-P)$ instead of matrix
$$
(A-P) = 
\left(
\begin{matrix}
	\alpha_{11} - P, & \alpha_{12},  & \ldots, & \alpha_{1n} \\
	\alpha_{21}, & \alpha_{22} - P,  & \ldots, & \alpha_{2n} \\
	\vdots & \vdots  & \ddots & \vdots \\
	\alpha_{n1}, & \alpha_{n2},  & \ldots, & \alpha_{nn} - P \\
\end{matrix}
\right).
$$
One of the many solutions \eqref{eq_eigenmat} is the solution to the equation
\begin{equation}\label{eigen_sys}
	\left(
	A-P
	\right)
	\left(
	\begin{matrix}
		\sigma_1 \\
		\sigma_2 \\
		\ldots \\
		\sigma_n \\
	\end{matrix}
	\right) = 0.
\end{equation}
Since all elements of the matrix $(A-P)$ commute with each other, we can consider the determinant of the matrix $(A-P)$, which must be equal to zero, since 
the coefficients $\{\sigma^\eta\}$ are in the nontrivial kernel of the matrix $(A-P)$. Hence the equation on the right functions of RLO arises
\begin{equation}
	\det\left(A - P\right) \equiv 0.
\end{equation}

The determinant is a polynomial of the operator $P$ of degree $n$, and its roots are various right-hand functions of RLO. 
Then, by substituting the obtained roots into the equation \eqref{eq_eigenmat}, we can find their corresponding coefficients 
$\{\sigma_\eta\}$ of the RLO's.

\section{Irreducible representations of the $su^J(2)$ algebra}
\subsection{Constructing the Casimir ladder operators}
In this section we construct ladder operators for the Casimir operator for the integer $s$.
In this case only irreducible representations with integer weights are realized, and the 
kernel of operator $J_z$ is always nontrivial for any number of particles. Thus, any irreducible 
representation can be obtained by applying the ladder operators of the algebra $su(2)$ to any state 
lying in the kernel of operator $J_z$. Thus it suffices to solve the classification problem within 
the kernel of $J_z$. When $s$ is half-integer, the irreducible representations of all possible weights are also realized. 
However, the proposed approach can be easily modified: resulting ladder operators may not commute
with the operator $J_z$ but they commute with ladder operators of $J_z$.

Let $s$ be a non-negative integer and consider two sets of operators $\{p^\dagger_k\}_{k=0}^s$ and 
$\{m^\dagger_k\}_{k=1}^s$ commuting with operator $J_z$
\begin{equation}
	\begin{matrix}
		p^\dagger_0 = 2a_0^{\dagger},\\
		\displaystyle p^\dagger_k = \frac{1}{\prod_{i=1}^{k}\sqrt{(s+i)(s-i+1)}}\left( a_{-k}^{\dagger} J_+^k + a_k^\dagger J_-^k \right),\\
		\displaystyle m^\dagger_k = \frac{1}{\prod_{i=1}^{k}\sqrt{(s+i)(s-i+1)}}\left( a_{-k}^{\dagger} J_+^k - a_k^\dagger J_-^k \right).
	\end{matrix}
\end{equation}
All operators from the sets $\{p^\dagger_k\}$ and $\{m^\dagger_k\}$ are ladder operators
of operator $N$
$$
[N, p^\dagger_k] = p^\dagger_k, \quad [N, m^\dagger_k] = m^\dagger_k.
$$
The operators $p_k^\dagger$ and $m_k^\dagger$ are closed with respect to the action of the Casimir operator $J^2$ in the sense of the definition \eqref{act_HTmu}
\begin{equation}\label{comm_Jp}
	\begin{matrix}
		[J^2, \, p^\dagger_0 ] = s(s+1) p^\dagger_0 + 2 s(s+1) p^\dagger_1,\\
		[J^2, \, p^\dagger_k ] = ((s+k+1)(s-k)-k(k-1))p^\dagger_k + (s+k+1)(s-k)p^\dagger_{k+1} +\\
		+ p_{k-1}^\dagger((\hat{j} + J_z + 1)(\hat{j} - J_z) - k(k-1))
		+ 2k (m^\dagger_k + m^\dagger_{k-1})J_z,\\
		[J^2, \, m^\dagger_k ] = ((s+k+1)(s-k)-k(k-1))m^\dagger_k + (s+k+1)(s-k)m^\dagger_{k+1} +\\
		+ m_{k-1}^\dagger((\hat{j} + J_z + 1)(\hat{j} - J_z) - k(k-1))
		+ 2k (p^\dagger_k + p^\dagger_{k-1})J_z,
	\end{matrix}
\end{equation}
where the operator $\hat{j}$ is defined as 
\begin{equation}\label{def_j}
	\hat{j} = \frac{1}{2} (\sqrt{\hat{I}+ 4J^2} - \hat{I}).
\end{equation}

Let us find the right-hand functions of the ladder operators from the equation \eqref{eq_eigenmat}.
We will construct ladder operators for the kernel $J_z$ since the whole basis of the irreducible 
representation can be restored by the action of operators $J_\pm$. For this reason we can replace 
the operator $J_z$ in the equation \eqref{comm_Jp} by zero $J_z = 0$ 
\begin{equation}\label{comm_Jp2}
	\begin{matrix}
		[J^2, \, p^\dagger_0 ] = s(s+1) p^\dagger_0 + 2 s(s+1) p^\dagger_1,\\
		[J^2, \, p^\dagger_k ] = ((s+k+1)(s-k)-k(k-1))p^\dagger_k + (s+k+1)(s-k)p^\dagger_{k+1} +\\
		+ p^\dagger_{k-1}((\hat{j} + 1)\hat{j} - k(k-1)),\\
		[J^2, \, m^\dagger_k ] = ((s+k+1)(s-k)-k(k-1))m^\dagger_k + (s+k+1)(s-k)m^\dagger_{k+1} +\\
		+ m^\dagger_{k-1}((\hat{j} + 1)\hat{j} - k(k-1)).
	\end{matrix}
\end{equation}
Let us construct a matrix $(A-P)$. The matrix $A$ is a block-diagonal
$$
A=\left(
\begin{matrix}
	P & 0\\
	0 & M
\end{matrix}
\right),
$$
consisting of two tridiagonal matrices ($P$ and $M$) of dimensions $s+1$ and $s$ correspondingly. Matrix 
$P$ has the following form:
\small

\begin{equation*}
	P = \left(
	\begin{matrix}
		s(s+1) & \hat{j}(\hat{j}+1),& 0, & \ldots & 0 & 0\\
		2s(s+1) & s(s+1)-4 & (\hat{j}-1)(\hat{j}+2) & \ldots & 0 & 0 \\
		0 & (s-1)(s+2) & s(s+1)-8  & \ldots & 0 & 0 \\
		\vdots & \vdots & \vdots & \ddots & \vdots & \vdots\\
		0 & 0 & 0 & \hdots &-s^2+5s-2 & \hat{j} (\hat{j} + 1) - s^2+s \\
		0 & 0 & 0 & \hdots & 2s & -s^2+s \\
	\end{matrix}
	\right),	
\end{equation*}

\normalsize
while $M$ is obtained from matrix $P$ by crossing out the first row and column. The choice of coefficients 
makes $P$ and $M$ symmetric with eigenvalues which are expressed through the operator $\hat{j}$. Ther following 
set of eigenvalues corresponds to matrix $A$
\begin{equation*}
	\{\theta (\theta + 2\hat{j} + 1)\hat{I} \}_{\theta = -s}^{s}.
\end{equation*}
The matrix $P$ is matched by $\theta$ of the same parity as $s$, and the matrix $M$ by all others.

We will look for the coefficients recurrently, starting from $\sigma_s$. For matrices $P$ and $M$ the equations 
on the ladder operator will be similar, which allows us to obtain a general result for them. 
Let us find the solution of the following equation 
\begin{equation*}
	\left(P-\theta (\theta + 2\hat{j} + 1)\hat{I}\right)
	\left(
	\begin{matrix}
		\sigma_0^\theta\\
		\sigma_1^\theta\\
		\vdots \\
		\sigma_s^\theta = \hat{I}\\
	\end{matrix}
	\right)
	= 0.		
\end{equation*}
The coefficient $\sigma_{s-1}^\theta$ may be found at $\sigma_s^\theta = \hat{I}$ and
\begin{equation}
	\sigma_{s-1}^\theta = \hat{j} \frac{\theta}{s} + \frac{\theta^2 + \theta + s^2 -s}{2s}.
\end{equation}
Then considering the $k$-th string
\begin{multline*}
    (s-k)(s+k+1)\sigma_{k-1}^\theta 
	+((s-k)(s+k+1)-k(k-1) - \theta(\theta + 2\hat{j} + 1))\sigma_{k}^\theta 
	+(\hat{j}-k)(\hat{j}+k+1)\sigma_{k+1}^\theta = 0,
\end{multline*}
and expressing $\sigma_{k-1}^\theta$ through  $\sigma_{k}^\theta$ and $\sigma_{k+1}^\theta$ 
we can obtain 
\begin{equation}
	\sigma_{k-1}^\theta = \frac{\theta^2 + \theta + k^2 + k}{(s+k)(s-k+1)} \sigma_{k}^\theta + 
	\hat{j} \frac{2\theta \sigma_{k}^\theta}{(s+k)(s-k+1)}-
	\sigma_{k}^\theta - 
	\frac{(\hat{j}+k+1)(\hat{j}-k)}{(s+k)(s-k+1)}\sigma_{k+1}^\theta.
\end{equation}
Here $\sigma_{k}^\theta$ is a polynomial of the operator $\hat{j}$ of degree $(s-k)$.

\subsection{Properties of Casimir ladder operators}

Let as denote the obtained ladder operators throught $\{\tau^\dagger_\theta\}_{\theta = - s}^{s}$:

\begin{equation}
	\begin{cases}
		\displaystyle \tau^\dagger_\theta = \sum_{k = 0}^s  p_k^\dagger \sigma_{k}^\theta , 
		& \text{for } \theta \text{ of the same parity as } $s$,\\
		\displaystyle \tau^\dagger_\theta = \sum_{k = 1}^s  m_k^\dagger \sigma_{k}^\theta , 
		& \text{otherwise.}
	\end{cases}
\end{equation}
Ladder operators have the following commutative relations with the $J^2$ operator
\begin{equation}
	[J^2, \, \tau^\dagger_\theta] = \tau^\dagger_\theta \theta (\theta + 2\hat{j} + 1).
\end{equation}
Since the Casimir operator $J^2$ is represented as a polynomial $J^2 = \hat{j}(\hat{j}+1)$ of operator $\hat{j}$,
we obtain commutation relations between  $\hat{j}$ and $\{\tau^\dagger_\theta\}$ from the solution of equation 
$$
	[\hat{j^2} +\hat{j}, \, \tau^\dagger_\theta] = \tau^\dagger_\theta\ X(X+2\hat{j} + 1)
	= \tau^\dagger_\theta\ \theta(\theta+2\hat{j} + 1).
$$
Hence $X = \theta \hat{I}$ and 
\begin{equation}
    [j, \, \tau^\dagger_\theta] = \theta \tau^\dagger_\theta.
\end{equation}
Operators $\{\tau^\dagger_\theta\}$ are also ladder operators for operators $\displaystyle \frac{1}{2\hat{j} + (2 k + 1)\hat{I}}$, where $k$ is non-negative:
\begin{equation}
	\left[\frac{1}{2\hat{j} + (2 k + 1) \hat{I}}, \, \tau^\dagger_\theta\right] = 
	\tau^\dagger_\theta \left(
	\frac{1}{2\hat{j} + (2 k + 1) \hat{I}} - \frac{1}{2\hat{j} + 
	(2 (k-\theta)+1) \hat{I}}
	\right).
\end{equation}
There is a similar expression with the left-hand function:
\begin{equation}
	\left[\frac{1}{2\hat{j} + (2k + 1) \hat{I}}, \, \tau^\dagger_\theta\right] = 
	\left(
	\frac{1}{2\hat{j} + (2k + 1) \hat{I}} - \frac{1}{2\hat{j} + (2(k-\theta)+1) \hat{I}}
	\right) \tau^\dagger_\theta.
\end{equation}
For an arbitrary polynomial of functions $\{\hat{j}^k\}_{k=0}^{n}$ and 
$\displaystyle \left\{\left(\frac{1}{2\hat{j} + (1 + 2 k) \hat{I}}\right)^k\right\}_{k=0}^{n}$
commutative relations with $\{\tau_\theta\}$ or $\{\tau^\dagger_\theta\}$ can be obtained. Single-particle Fock states belongs to the irreducible representation of the algebra $su(2)$ corresponding to the eigenvalue $s(s+1)$ of the Casimir operator $J^2$
\begin{equation*}
	J^2 \left|0, \, 0, \, \ldots, \, n_k = 1, \, \ldots , 0 \right> = 
	s(s+1)  \left|0, \, 0, \, \ldots, \, n_k = 1, \, \ldots , 0 \right>.	
\end{equation*}
The kernel $J_z$ is one-dimensional and spaned by the following vector
\begin{equation*}
	\left|0, \, 0, \, \ldots, \, n_0 = 1, \, \ldots , 0 \right>.	
\end{equation*}
The action of the ladder operators $\{\tau^\dagger_\theta\}$ allows constructing the canonical basis 
of the kernel $J_z$. At these, it is easy to show that
\begin{equation*}
	[\tau^\dagger_\theta \tau_\theta, \, J^2] = 0 = [\tau^\dagger_\theta \tau_\theta, \, J_z] = [\tau^\dagger_\theta \tau_\theta, \, N].	
\end{equation*}
Thus, the commuting set $\left\{N; J^2, j_z\right\}$ can be augmented to complete set of commuting operators 
by some self-adjoint polynomials of ladder operators.

\subsection{Annihilated states of the Casimir ladder operators}

The geometry of Fock space allows us to find annihilated states of Casimir ladder operators 
$\{\tau^\dagger_\theta, \, \tau_\theta\}_{\theta = - s}^{s}$. Consider the eigenvectors of the operators 
$\hat{j}$ and $N$ lying in the kernel of the operator $J_z$
\begin{equation*}
	\left|n, \, j, \, , j_z = 0\right>,	
\end{equation*}
Given $n$ the eigenvalues of the operator $\hat{j}$ are placed on the interval $0 \leqslant j \leqslant ns$. 
The action of operators within the $J_z$ kernel can be represented by the following scheme for $\omega = 1\ldots s$:
\begin{equation}
	\begin{matrix}
		&\tau^\dagger_\omega: \quad \left|n, \, j \right> 
		\mapsto \left|n + 1, \, j + \omega \right>,  \\
		&\tau_\omega: \quad \left|n + 1, \, j + \omega \right> 
		\mapsto \left|n, \, j \right>, 
	\end{matrix}
		\\ 	\text{and} \\
	\begin{matrix}
		&\tau^\dagger_{-\omega}: \quad  
		\left|n, \, j + \omega \right> 
		\mapsto \left|n + 1, \, j  \right>, \\
		&\tau_{-\omega}: \quad  
		\left|n + 1, \, j \right> 
		\mapsto \left|n, \, j + \omega \right>. \\
	\end{matrix}
\end{equation}

Operators $\tau^\dagger_\omega$ have a trivial kernel if $\omega$ is the same parity, as $s$. If $\omega$ differs in parity from $s$, 
then all one-particle state lies in the kernel of $\tau^\dagger_\omega$. This is due to the antisymmetric definition of the operator 
$\tau^\dagger_\omega$ for $\omega$ other than $s$ parity.

The operators $\tau_\omega$ will annihilate all states $j < \omega$ and the vacuum state $n = 0$, thus realizing $\omega$ different 
representations of the algebra. The algebra of the pair of operators $\tau^\dagger_{\omega}$ and $\tau_{\omega}$ itself is a 
deformation of the Weyl algebra $w(1)$. Its different representations are defined by the number $r_\theta = j \mod \omega$ and 
the eigenvalues of of the self-adjoint operators 
\begin{equation*}
	\tau^\dagger_\omega \tau_\omega.	
\end{equation*}

The operators $\tau^\dagger_{-\omega}$ annihilate all states with $j < \omega$, while the operators $\tau_{-\omega}$ annihilate all 
states with $j > ns - \omega$ and vacuum state $n = 0$. Thus, we can say that the operators $\tau^\dagger_{-\omega}$ and $\tau_{-\omega}$ 
represent a deformation of the algebra $su(2)$, where the representations may be classified by the number $r_\theta = j \mod \omega$ 
and the eigenvalues of the following selfadjoint operators
\begin{equation*}
	L_z^\omega = [\tau^\dagger_{-\omega}, \tau_{-\omega}], \quad
	L^2_\omega = (L_z^\omega)^2 + \frac{1}{2}\left(
	\tau^\dagger_{-\omega} \tau_{-\omega} + \tau_{-\omega} \tau^\dagger_{-\omega}
	\right).	
\end{equation*}
The operators $\tau^\dagger_0$ and $\tau_0$ do not change the eigenvalues of the Casimir operator $J^2$
\begin{equation}
	\begin{matrix}
		&\tau^\dagger_0 \left|n, \, j \right> 
		\Rightarrow \left|n + 1, \, j \right>,  \quad
		&\tau_0 \left|n + 1, \, j, \right> 
		\Rightarrow \left|n, \, j \right>, 
	\end{matrix}
\end{equation}

\section{Demo case: $s = 1$}
In this case the classification problem is of small interest because of all subspaces of kernel $J_z$ are one-dimensional 
and the set of commuting operators $N; \, J^2,\,J_z$ is complete. However, the use of ladder operators can be well demonstrated 
by this example. In this case the generators of the $su(2)$ algebra are represented as follows
\begin{equation}\label{def_su2_Jordan}
	\begin{matrix}
		J_z = \sum\limits_{\mu = -1}^{1} \mu a_{\mu}^{\dagger} a_{\mu},\quad
		J_+ = (J_-)^{\dagger} = \sum\limits_{\mu = -1}^{\mu = 0} \sqrt{(\mu + 2)(1 - \mu)} a_{\mu + 1}^{\dagger} a_ {\mu}.
	\end{matrix}
\end{equation}
Then we consider the operators 
\begin{equation}\label{def_p0p1}
	p_0^{\dagger} = 2a_0^{\dagger}, \quad
	\sqrt{2} p_1^{\dagger} = a_1^{\dagger} J_- + a_{-1}^{\dagger} J_+,
\end{equation}
with commutation relations
\begin{equation}
	\begin{matrix}
		[p_0, \, p_0^\dagger] = 4, \quad
		[p_1, \, p_1^\dagger] = 2\hat{j}(\hat{j} + 1) - J_z(2 J_z + 1) + (N-N_0)(J_z-2), \\
		[p_1, \, p_0^\dagger] = 2(N-N_0), \quad
		[p_0, \, p_1^\dagger] = 2(N-N_0),
	\end{matrix}
\end{equation} 
where $N_0 = a^\dagger_0 a_0$.
Considering the action of the $m_1^\dagger$ operator on an arbitrary Fock state 
\begin{equation*}
	\left| n_{-1} = m, \, n_0 = k, \, n_{1} = m \right>	
\end{equation*}
one can easily check that the operator $m_1^\dagger$ annihilates the $J_z$ kernel. However, outside the kernel $J_z$ the operator 
$m_1^\dagger$ acts nontrivially, which allows the construction of ladder operators on the whole Fock space. Our goal is to obtain 
canonical basis inside the kernel of $J_z$ and reconstruct whole basis applying the operators $J_+$ and $J_-$.

Let us write commutation relations between $p_i^\dagger$ and $J^2$ operators:
\begin{equation*}
	\begin{matrix}
		[J^2, \, p_0^\dagger] = 2 p_0^\dagger + 4 p_1^\dagger,\quad
		[J^2, \, p_1^\dagger] = p_0^\dagger J_-J_+ = p_0^\dagger (\hat{j} - J_z)(\hat{j} + J_z + 1).
	\end{matrix}	
\end{equation*}
Taking $J_z \equiv 0$ we can rewrite this relations as follows 
\begin{equation}
	\begin{matrix}
		[J^2, \, p_0^\dagger] = 2 p_0^\dagger + 4 p_1^\dagger,\quad
		[J^2, \, p_1^\dagger] = p_0^\dagger \hat{j} (\hat{j} + 1).
	\end{matrix}
\end{equation}
The solution of the equation on the right functions of the ladder operators are operators
$-2\hat{j}$ and $2(\hat{j} + 1)$.

We introduce the notations $\tau^\dagger_{-1}$ and $\tau^\dagger_{1}$ for the obtained ladder operators.
They have the following commutative relations with the operator $J^2$
\begin{equation}
	\begin{matrix}
		[J^2, \, \tau^\dagger_{-1}] = -\tau^\dagger_{-1} 2 \hat{j},\quad
		[J^2, \, \tau^\dagger_{ 1}] = \tau^\dagger_{ 1}  2 (\hat{j} + 1)\\
	\end{matrix}
\end{equation}
and may be expressed through the operators $p_0^\dagger$ and $p_1^\dagger$ as follows
\begin{equation}
	\begin{matrix}
		\tau^\dagger_{-1} = p_0^\dagger \hat{j} - 2 p_1^\dagger
		,\quad
		\tau^\dagger_{ 1} = p_0^\dagger (\hat{j} + 1) + 2 p_1^\dagger.
	\end{matrix}
\end{equation}
Commutation relations between operators $\tau^\dagger_{-1}$ $\tau^\dagger_{1}$ and $\hat{j}$ gives
\begin{equation}
	\begin{matrix}
		[\hat{j}, \, \tau^\dagger_{-1}] = -\tau^\dagger_{-1},\quad
		[\hat{j}, \, \tau^\dagger_{ 1}] = \tau^\dagger_{ 1}\\
	\end{matrix}
\end{equation}
From the Jacobi relation we can also obtain commutator $[\tau^\dagger_{1} ,\, \tau^\dagger_{-1}]$ that is ought 
to be a ladder operator $\hat{j}$:
\begin{equation*}
	[\hat{j}, \, [\tau^\dagger_{1} ,\, \tau^\dagger_{-1}]] = 2 [\tau^\dagger_{1} ,\, \tau^\dagger_{-1}].
\end{equation*}
Any vector of the canonical basis can be obtained by the joint action of the ladder operators
\begin{equation*}
     \left|n, \, j, \, j_z \right>_{su2} = \alpha(n,j,j_z) 
     \begin{cases}
        J_+^{j_z}  
        (\tau^\dagger_{-1})^{\frac{n-j}{2}}
        (\tau^\dagger_{1})^{\frac{n+j}{2}}|000\rangle_{F}, & j_z>0 \\
        (\tau^\dagger_{-1})^{\frac{n-j}{2}}
        (\tau^\dagger_{1})^{\frac{n+j}{2}}|000\rangle_{F}, & j_z=0 \\
        J_-^{j_z}
        (\tau^\dagger_{-1})^{\frac{n-j}{2}}
        (\tau^\dagger_{1})^{\frac{n+j}{2}}|000\rangle_{F}, & j_z<0.
     \end{cases}
\end{equation*}
The action of the ladder operators and the structure of irreducible representations of the algebra $su(2)$ can be 
visualized by the following scheme for $j_z = 0$:
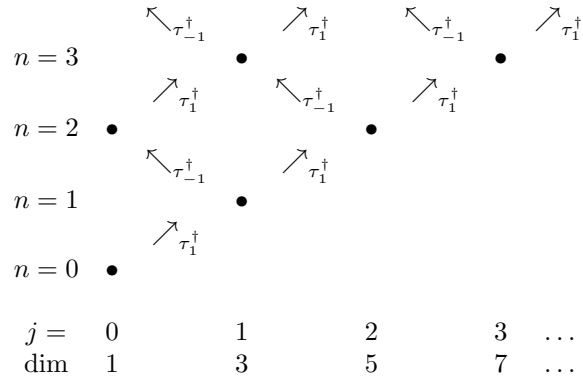
\begin{equation*}
	\begin{matrix}
		{\quad} & {} & {{\nwarrow }_{{{\tau^\dagger_{-1}}}}} & 
		{} & {{\nearrow }_{{{\tau^\dagger_{1}}}}} & {} & 
		{{\nwarrow }_{{\tau^\dagger_{-1}}}} & {} & {{\nearrow }_{{\tau^\dagger_{1}}}} \\
		   n=3 & {} & {} & 
		   \bullet  & {} & {} & 
		   {} & \bullet  & {}   \\
		   {} & {} & {{\nearrow }_{{{\tau^\dagger_{1}}}}} & 
		   {} & {{\nwarrow }_{{{\tau^\dagger_{-1}}}}} & {} & 
		   {{\nearrow }_{{\tau^\dagger_{1}}}} & {} & {}  \\
		   n=2 & \bullet  & {} & 
		   {} & {} & \bullet  & 
		   {} & {} & {}   \\
		   {} & {} & {{\nwarrow }_{{{\tau^\dagger_{-1}}}}} & 
		   {} & {{\nearrow }_{{{\tau^\dagger_{1}}}}} & {} & 
		   {} & {} & {}    \\
		   n=1  & {}        & {} & \bullet & {} & {}& {} & {}  & {}   \\
		   {}   & {}        & {{\nearrow }_{{{\tau^\dagger_{1}}}}} & {} & {} & {} & {} & {}  & {}  \\
		   n=0  & \bullet   & {} & {}   & {} & {}   & {} & {}  & {} \\
		   {}   & {}        & {} & {}   & {} & {}   & {} &  {} & {} \\
		   j=   & 0         & {} & 1    & {} & 2    & {} &  3  & {\ldots} \\
		   \dim & 1         & {} & 3    & {} & 5    & {} &  7  & {\ldots} \\
		\end{matrix}	
\end{equation*}
Now let us consider again the operators $p_0^\dagger$ and $p_1^\dagger$ which may be expressed through the 
operators $\tau^\dagger_{-1}$ and $\tau^\dagger_{1}$:
\begin{equation*}
	p_0^\dagger = \left( \tau^\dagger_{1} + \tau^\dagger_{-1} \right) \frac{1}{2\hat{j} + 1}, \quad
	p_1^\dagger = \frac{1}{4} \left( (\tau^\dagger_{1} - \tau^\dagger_{-1})- (\tau^\dagger_{1} + \tau^\dagger_{-1}) \frac{1}{2\hat{j} + 1}\right),
\end{equation*}
From here we can find commutation relations with the operator $\hat{j}$: 
\begin{equation}\label{comm_jp5}
	[\hat{j}, \, p_0^\dagger] = (p_0^\dagger + 4 p_1^\dagger) \frac{1}{2\hat{j} + 1},
	\quad
	[\hat{j}, \, p_1^\dagger] = (p_0^\dagger J^2 - p_1^\dagger) \frac{1}{2\hat{j} + 1}.
\end{equation}
We define a new operators, which will also be RLOs for the operator $J^2$:
\begin{equation}
	A^\dagger = \tau^\dagger_1 \frac{1}{2\sqrt{\hat{j}+1}} \frac{1}{\sqrt{(N+1) + \hat{j}+1 + 1}} 
	\frac{\sqrt{2(\hat{j}+1) +1}}{\sqrt{2(\hat{j}+1) - 1}},
\end{equation}
\begin{equation}
	L_+ = \tau^\dagger_{-1} \frac{1}{2\sqrt{2}\sqrt{\hat{j}+1}}  
	\frac{\sqrt{2(\hat{j}+1) -1}}{\sqrt{2(\hat{j}+1) + 1}}.
\end{equation}
Operators $A$ and $A^\dagger$ satisfy the commutation relations on the Weyl algebra $w(1)$
\begin{equation*}
	[A, \, A^\dagger] = \hat{I}.
\end{equation*}
Self-adjoint operator $A^\dagger A$ has the same eigenvalues as the operator $\hat{j}$. 
The action on the state $\left| n, \, j \, j_z \right>$ is defined by the formula 
\begin{equation*}
	A^\dagger A \left| n, \, j \, j_z \right> = j \left| n, \, j \, j_z \right>.	
\end{equation*}
Operators $L_\pm$ are represented by self-adjoint polynomials and define the operators $L_z$ and $L^2$ 
\begin{equation*}
	L_z = \frac{1}{2}[L_+, \, L_-], \quad L^2  = L_z^2 + L_z + L_- L_+,	
\end{equation*}
which satisfy the commutation relations on $su(2)$ algebra. Action $L_z$ and $L^2$ on the eigenstates 
is given by the following expression
\begin{gather*}
	L_z \left| n, \, j \, j_z \right> = \left( \frac{n-j}{2} - \frac{n+j}{4} \right)\left| n, \, j \, j_z \right>,	\\ 
	L^2 \left| n, \, j \, j_z \right> = \frac{n+j}{4} \left(\frac{n+j}{4} + 1\right)\left| n, \, j \, j_z \right>.
\end{gather*}
The set of mutually commutative operators $\left\{A^\dagger A, L_z, L^2\right\}$ is complete and can be used to classify 
the states as well as the sets $\left\{N_{-1}, N_{0}, N_{1}\right\}$ and $\left\{J^2, J_z, N\right\}$. By constructing 
left-hand ladder operators for $J^2$, we obtain another form of ladder operators
\begin{equation}
	\begin{matrix}
			\bar{\tau}_1 = [a_0, \hat{j}] + a_0, & \bar{\tau}_{-1} = -[a_0, \hat{j}] + a_0,\\
			\bar{\tau}^\dagger_1 = [\hat{j}, a_0^\dagger] + a_0^\dagger, & \bar{\tau}_{-1} = -[\hat{j}, a_0^\dagger] + a_0^\dagger,\\
	\end{matrix}
\end{equation}
from where, in particular, an interesting expression emerges
\begin{equation*}
	[\hat{j}, [\hat{j},\, a_0^\dagger]] = a_0^\dagger.
\end{equation*}

\section{Conclusion}
A method of classification and construction of invariant spaces corresponding to various irreducible representations of 
the $su(2)$ algebra is proposed. For Casimir operator of this algebra we obtained a set of ladder operators, which are used 
to find the canonical basis. Algebras formed by ladder operators are deformations of known algebras, which eigenvalues 
determine persistent states of the Hamiltonian $H$. In this paper we considered the simplest case for the $su^j(2)$ representation 
and applied the ladder operator approach to demonstrate the method. This approach is based on commutative algebra relations 
and can be applied to the analysis of irreducible representations of various Lie algebras. Having solved the problem in this 
paper we obtained an infinite basis of a complex structure which can be recovered from any chosen element of basis by the 
action of the ladder operators.

\end{document}